\documentclass[twocolumn,showpacs,preprintnumbers,amsmath,amssymb,showkeys]{revtex4}

\usepackage{graphicx}
\usepackage{dcolumn}
\usepackage{bm}
\usepackage{color}

\begin{document}

\preprint{Superlattices and Microstructures, Volume 29, Issue 2, February 2001, Pages 99-104}

\title{GaAs $\delta$-doped quantum wire superlattice characterization by quantum Hall effect and Shubnikov-de Haas oscillations}

\author{T. Ferrus \footnote{Present address : Hitachi Cambridge Laboratory, J. J. Thomson Avenue, CB3 0HE, Cambridge, United Kingdom}}
\email{taf25@cam.ac.uk}
\author{B. Goutiers}
\author{J. Galibert}
\author{L. Ressier}
\author{J. P. Peyrade}

\affiliation {Institut National des Sciences appliqu\'ees, 135 Avenue de Rangueil, 31077 Toulouse, France}

\date{\today}

\begin{abstract}

Quantum wire superlattices (1D) realized by controlled dislocation slipping in quantum well superlattices (2D) (atomic saw method) have already shown Magneto-phonon oscillations. This effect has been used to investigate the electronic properties of such systems and prove the quantum character of the physical properties of the wires. By cooling the temperature and using pulsed magnetic field up to 35 T, we have observed both quantum Hall effect (QHE) and Shubnikov–de Haas (SdH) oscillations for various configurations of the magnetic field. The effective masses deduced from the values of the fundamental fields are coherent with those obtained with Magneto-phonon effect. The field rotation induces a change in the resonance frequencies due to the modification of the mass tensor as in a (3D) electron gas. In view the QHE, the plateaus observed in $\rho_{yz}$ are dephased relatively to $\rho_{zz}$ minima which seems to be linked to the dephasing of the minima of the density of states of the broadened Landau levels.

\end{abstract}

\maketitle

Full article available at doi:10.1006/spmi.2000.0928

\end{document}